\documentclass[journal]{IEEEtran}
\usepackage{amsmath,amssymb,amsfonts}
\usepackage{graphicx}
\usepackage{textcomp}
\usepackage{xcolor}
\usepackage{epsfig}
\usepackage{multirow}
\usepackage{float}
\usepackage[ruled,vlined,linesnumbered]{algorithm2e}
\usepackage[caption=false]{subfig}
\usepackage{varwidth}
\usepackage{url}
\usepackage{amssymb}
\usepackage{makecell}
\usepackage{amsthm}
\usepackage{mathtools}
\usepackage{hyperref}
\usepackage{soul,color}

\begin{document}

\title{A Survey on Blockchain for Big Data: \\Approaches, Opportunities, and Future Directions}

\author{Deepa~N, Quoc-Viet~Pham, Dinh~C.~Nguyen, Sweta~Bhattacharya, B.~Prabadevi, Thippa~Reddy~Gadekallu, Praveen~Kumar~Reddy~Maddikunta, Fang~Fang, Pubudu~N.~Pathirana

\thanks{Deepa~N, Sweta~Bhattacharya, B.~Prabadevi, Thippa~Reddy~Gadekallu, Praveen~Kumar~Reddy~Maddikunta are with the School of Information Technology and Engineering, Vellore Institute of Technology, Vellore 632014, India (e-mail: \{deepa.rajesh, sweta.b, prabadevi.b, thippareddy.g, praveenkumarreddy\}@vit.ac.in).}

\thanks{Quoc-Viet Pham is with the Research Institute of Computer, Information and Communication, Pusan National University, Busan 46241, Korea (e-mail: vietpq@pusan.ac.kr).}

\thanks{Dinh~C.~Nguyen, Pubudu~N.~Pathirana are with School of Engineering, Deakin University, Waurn Ponds, Australia (email: cdnguyen@deakin.edu.au, pubudu.pathirana@deakin.edu.au)}

\thanks{Fang~Fang is with Department of Engineering, Durham University, Durham DH1 3LE, UK (email: fang.fang@durham.ac.uk)}

\thanks{This work was supported by a National Research Foundation of Korea (NRF) Grant funded by the Korean Government (MSIT) under Grants NRF-2019R1C1C1006143 and NRF-2019R1I1A3A01060518. Quoc-Viet Pham is the corresponding author.}
}% <-this % stops a space

\maketitle

\begin{abstract}
Big data has generated strong interest in various scientific and engineering domains over the last few years. Despite many advantages and applications, there are many challenges in big data to be tackled for better quality of service, e.g., big data analytics, big data management, and big data privacy and security. Blockchain with its decentralization and security nature has the great potential to improve big data services and applications. In this article, we provide a comprehensive survey on blockchain for big data, focusing on up-to-date approaches, opportunities, and future directions. First, we present a brief overview of blockchain and big data as well as the motivation behind their integration. Next, we survey various blockchain services for big data, including blockchain for secure big data acquisition, data storage, data analytics, and data privacy preservation. Then, we review the state-of-the-art studies on the use of blockchain for big data applications in different vertical domains such as smart city, smart healthcare, smart transportation, and smart grid. For a better understanding, some representative blockchain-big data projects are also presented and analyzed. Finally, challenges and future directions are discussed to further drive research in this promising area.
\end{abstract}


\IEEEpeerreviewmaketitle
\section{{Introduction}}
\label{Sec:Introduction}
The global data traffic has increased at an unprecedented rate over the last decade, thus the special interest in "big data". As reported in \cite{BigData_Market}, the big data market shall reach 229.4 billion \$ in 2025 and significantly reduce the expenditure for various vertical industries like healthcare, retail, transportation and logistics, manufacturing, media and entertainment.  
%
% According to a Cisco report \cite{}, due to the massiveness of the Internet of Things (IoT) devices and the emergence of applications, 
%
Despite the lack of a precise definition, attention to big data can be seen in many scientific and engineering areas, e.g., computer vision, Internet of Things (IoT) data analytics, operation management, and smart cities. Adding to the structural embodiment, \cite{hu2014toward} considered big data from three aspects, including attributive, comparative, and architectural. According to \cite{gantz2011extracting}, big data can be identified as a new generation of technologies and architectures investigated to analyze a large amount of data and capture its main characteristics (e.g., high velocity, knowledge discovery, and analytics). The comparative aspect considers big data as the datasets, which has a very large size and dimensionality and cannot be stored, managed, analyzed, and captured by conventional database tools \cite{manyika2011big}. From the architectural viewpoint, big data is identified as the datasets, which have very large volume, velocity, and representation, and require significant horizontal scaling methods for efficient processing \cite{pouyanfar2018multimedia}. 

Nevertheless, there are various challenges and issues associated with big data techniques and applications, for example, data security and privacy, energy management, scalability of computing infrastructure, data management, data interpretation, real-time data processing, big data intelligence. Among these challenges, security and privacy have been considered as important issues \textcolor{black}{since big data often involves different types of sensitive personal information, e.g., age, addresses, personal preference, banking details, etc.} There have been various solutions and techniques investigated to preserve data confidentiality and private information. An example is \cite{su2017security}, where matching theory and a coalitional game were jointly utilized to optimize a resource allocation problem so as to secure mobile social networks with big data. The use of reinforcement learning was investigated in \cite{wu2018big} to design a security-aware algorithm for a smart grid system. \textcolor{black}{Recently, blockchain as a ledger technology has emerged as attractive solutions for providing security and privacy in big data systems.}
%Blockchain, a disruptive technology to solve security and privacy issues in IoT, has also found vital applications in enhancing the security and privacy level of big data systems.
For example, it was shown in \cite{liu2019blockchain} that blockchain can play a vital role in providing high-quality data and securing data sharing for industrial IoT applications. 
In \cite{liu2020b4sdc}, a blockchain-based mechanism was proposed for securing data collection in mobile ad hoc networks and incentivizing mobile nodes for efficient data collection. 
Furthermore, blockchain was also integrated with edge computing servers to enhance the data quality and process the compute-intensive tasks requested by IoT devices with security guarantees \cite{xu2020become}. 
\textcolor{black}{With its unique advantages, blockchain has the great potential to transform current big data systems by providing efficient security features and network management capabilities for enabling newly emerging big data services and applications.}
%Obviously, blockchain has the great potential for enhancing big data infrastructures, services, and applications. 
% 
In this survey, we present a comprehensive review of blockchain for big data, ranging from approaches to opportunities and future directions. 

\subsection{{State of the Arts and Our Contributions}}
Due to the importance of blockchain and big data, there have been a number of surveys published in related topics over the past few years. One of the earliest surveys on blockchain was carried out in \cite{zheng2018blockchain}. Privacy and security issues of blockchain systems were reviewed in \cite{li2020survey, feng2019survey, sengupta2020comprehensive, saad2020exploring}. The survey in \cite{liu2019survey} presented applications (e.g., game for mining management, game for security/privacy issues, and game for blockchain applications) of game theories for blockchain systems. Various surveys have been conducted to study applications of blockchain for other technologies. For example, the possibility of utilizing blockchain for IoT systems can be found in \cite{reyna2018blockchain, dai2019blockchain, ali2018applications, sengupta2020comprehensive}. The integration of blockchain with edge computing and 5G systems were studied in \cite{yang2019integrated} and \cite{nguyen2020blockchain}, respectively. The surveys in \cite{zhuang2020blockchain, mollah2020blockchain} carried out reviews of applications and opportunities of blockchain for smart grid networks. Moreover, several surveys have been dedicated to reviewing the fundamentals and applications of big data analytics. A survey on techniques and technologies for big data management was presented in \cite{siddiqa2016survey}. Recent studies in \cite{ge2018big, mohammadi2018deep} reviewed and discussed the roles and applications of big data for IoT systems and smart cities. Big data analytics have also found applications in smart grid and intelligent transportation systems, and representative surveys can be found in \cite{ghorbanian2019big, zhu2019big}. The concept of mobile big data was reviewed in \cite{cheng2017mobile} and recently found many applications for next-generation wireless systems (e.g., 5G, beyond 5G, and 6G), from the physical and MAC layers to the application layer \cite{zhang2019deep, al2020survey}.

In spite of many research efforts, we are not aware of any survey that comprehensively studies the applicability of blockchain for big data applications. Although the survey in \cite{karafiloski2017blockchain} reviews blockchain for big data applications and challenges, it is very short and not updated since it has been published several years ago. The survey in \cite{tariq2019security} mainly reviews the use of blockchain to address security issues in edge computing-based IoT applications. Other surveys in \cite{nguyen2020blockchain, dai2019blockchain, nguyen2019integration} also mention the interplay between blockchain and big data, but they only provide brief introductions on this topic without an in-depth survey unlike our paper. Motivated by the above observations, we provide a comprehensive survey on blockchain for big data, which covers fundamental knowledge, up-to-date approaches, opportunities, research challenges, issues, and future directions. The key objective of this survey is to inspect the state-of-the-art studies and to carry out a review on the applicability of blockchain for big data applications. In summary, the contributions and features offered by this work can be stated as the following.
\begin{itemize}
    \item Firstly, we present an overview of blockchain and big data as well as \textcolor{black}{the motivations behind the use of blockchain for big data.} We show that blockchain has the great potential for facilitating big data analytics such as control of dirty data, enhanced security and privacy, enhanced quality of data, and the management of data sharing. 
    
    \item Secondly, we review four main blockchain services for big data, including blockchain for secure data acquisition, blockchain for secure data storage, blockchain for data analytics, and blockchain for data privacy preservation. 
    
    \item \textcolor{black}{Thirdly, we provide an extensive discussion of the use of blockchain in several popular big data applications, including smart healthcare, transportation and logistics, smart grid, and smart cities. Moreover, some popular blockchain-based big data projects are also introduced and analyzed.}
    %Thirdly, we carry out a review of blockchain applications and the use of blockchain on various vertical domains, for example, smart healthcare, transportation and logistics, smart grid, and smart cities. For a better understanding and representation, we also draw some examples of practical projects on blockchain for big data. 
    
    \item Finally, we discuss a number of research challenges that arose from the state-of-the-art survey on the use of blockchain for big data. We also highlight open research opportunities that provide a roadmap for future research. 
\end{itemize}

\subsection{{The Survey Organization}}
\begin{figure}[h!]
	\centering
	\includegraphics[width=0.8\linewidth]{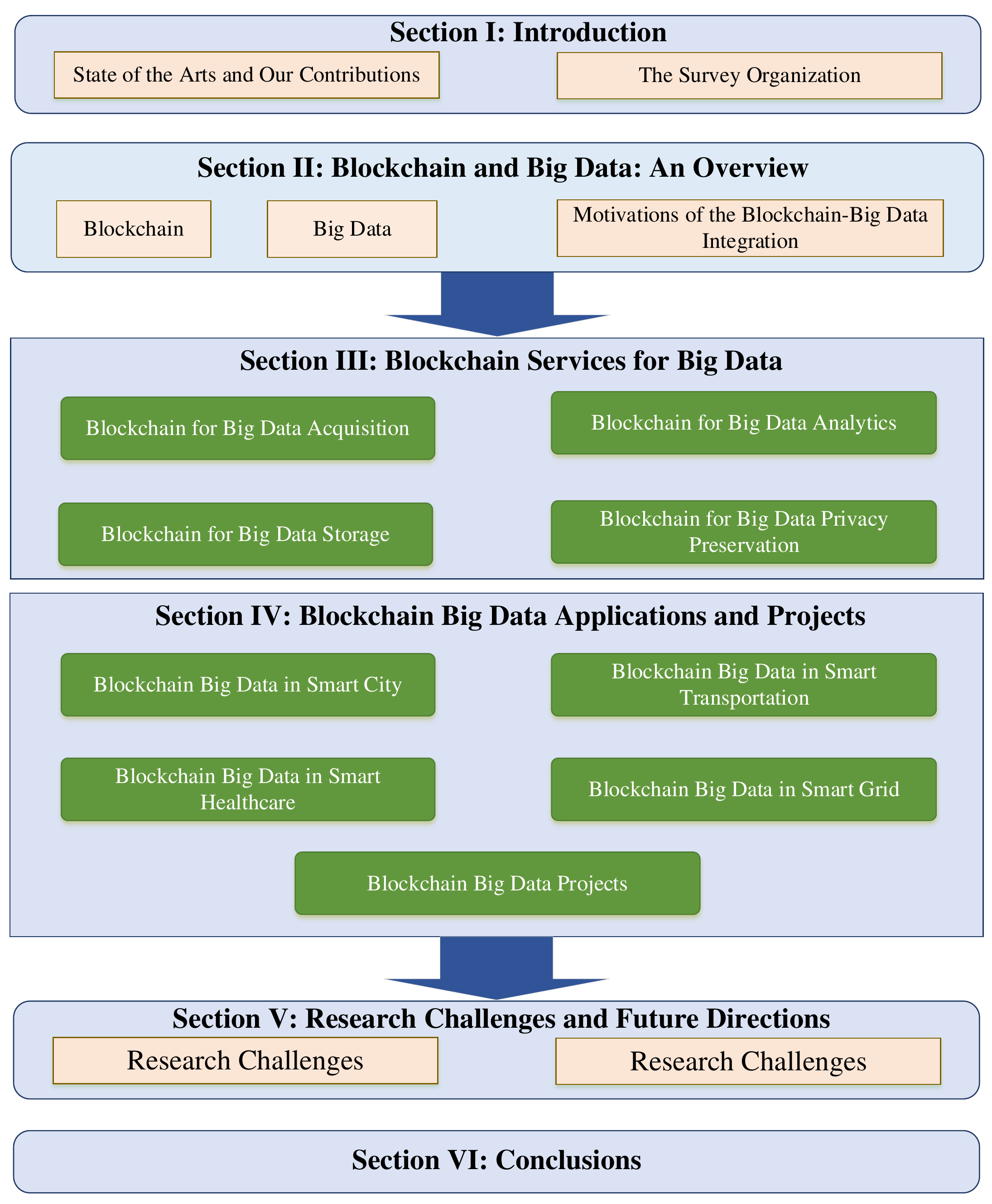}
	\caption{Organization of this article.}
	\label{Fig:PaperStructure}
\end{figure}

The structure of this survey is organized as Fig.~\ref{Fig:PaperStructure}. An overview of blockchain and big data is presented in Section~\ref{Sec:Overview}, along with a discussion of the motivations of their integration. The main parts of this survey are given in Sections~\ref{Sec:Blockchain_Services_BD} and~\ref{Sec:BlockchainBD_Applications}, which respectively present 1) blockchain services for big data and 2) blockchain-big data applications and projects. Section~\ref{Sec:Challenges_Future-Directions} discusses and highlights a number of research challenges, issues, and future directions. Finally, Section~\ref{Sec:Conclusion} concludes the article. 

\section{{Blockchain and Big Data: An Overview}}
\label{Sec:Overview}
This section presents the background and recent developments of blockchain, big data, and motivations of their integration.

\subsection{{Blockchain}}
Blockchain is presently one of the most prevalent disruptive technologies which is paving the way for emerging financial and industrial services \cite{zhou2020solutions,singh2020blockiotintelligence}.
%which is basically a method of storing information publicly and distributing the same.
Conceptually, it consists of a list of records commonly known as blocks wherein information stored is encrypted ensuring privacy and security. Also, unlike other technologies, \textcolor{black}{blockchain is a decentralized network wherein the participating members have complete authority to monitor all transactions in the blockchain network in a peer-to-peer (P2P) manner \cite{viriyasitavat2019blockchain,da2019application}.}  The blockchain technology is an amalgamation of varied multidisciplinary concepts such as software engineering, cryptography, distributed computing, creating an infrastructure emphasizing on digital assets related security.
\textcolor{black}{The combination of all these concepts is commonly termed as cryptoeconomics that create robust P2P networks for facilitating the use and transfer of assets among computers in digital markets \cite{berg2019understanding}.}
%The combination of all these concepts could be commonly termed as cryptoeconomics that help in the production, use and transfer of assets using computer network, cryptology and related concepts for the overall prosperity of digital markets \cite{berg2019understanding}. 
Cryptocurrency can be considered as the present and future mode of financial transactions which support the aforementioned transparency and security aspects of blockchain technology. In the digital market, cryptocurrencies are existing in different forms such as Bitcoin, Ethereum, Litecoin, Stellar, Ripple, Z-cash, Dash, etc \cite{yuan2018blockchain, abou2019blockchain}. 

\textcolor{black}{Bitcoin as the most popular cryptocurrency platform was introduced} by Sakato Nakamoto and since then almost 1600 cryptocurrencies have evolved using the Bitcoin concept \cite{shalini2019survey}. In the case of Bitcoin, whenever a sender initiates a transaction, it is sent to the receiver through the transaction being performed on the public bitcoin network. User verification is conducted by miners in the network who also ensure that the sender has the necessary number of bitcoins to be sent to the receiver without affecting the basic sanity of the network. After approval and verification by the miner, the transaction is added to the block which eventually becomes a part of the blockchain network. Finally, the relevant transactions pertaining to the block get executed thereby updating the ledgers across all the nodes so that all participants share the same copy of transaction for ensuring transparency and security \cite{parkin2019senatorial}. 

\textcolor{black}{Blockchain platforms can be classified into three types; public, private and hybrid blockchains, based on their areas of application \cite{casino2019systematic}.} A public blockchain does not have any specific single owner and are visible to everyone in the network. Bitcoin is an example of public blockchain which is decentralized with its consensus process being available to all participants in the network. The private blockchain on the contrary is permissioned and controls the participation of network members to read from and write to the blockchain. In hybrid blockchains, the public access is given to only specific group. It is a partially decentralized framework where the consensus process is guided by rules agreed among all parties regarding the control and access over the blockchain \cite{zhang2020blockchain}. Some of the most important features of blockchain are as follows: 
\begin{enumerate}
\item \textit{Immutability}: Blockchain is almost impossible to corrupt due to a permanent and unalterable network. It works differently from the traditional banking system using collection of nodes and each node in the system has a copy of the digital ledger \cite{huang2020survey}. When any transaction is initiated, nodes check its validity and authenticate to add to the ledger. Hence, the success of any transaction depends on the consensus across all major nodes which makes the framework transparent and secure. It also eliminates the chances of corruption which is especially evident in a public blockchain that allows everyone to see the transactions but does not allow altering the data stored in the blockchain \cite{dey2020role, angelis2019blockchain}. 

\item \textit{Decentralization}: The network is not governed by a single authority but a group of nodes that are responsible for maintaining the network. This decentralized approach allows participants to access the blockchain from the web \textcolor{black}{and store their replicated information using private keys \cite{chen2020blockchain}.}

\item \textit{Security}: Blockchain with its decentralized and immutable natures can provide high degrees of security \cite{zhou2020solutions,moin2019securing}. The use of cryptography includes implementation of complex algorithms acting as firewalls against unauthorized attacks. Each information is hashed which hides its actual nature and also provides a unique identification for each data. In the chain, each block in the ledger holds its own hash and also the hash of its previous block which makes it immutable to tamper the data. Hashing also makes the framework irreversible. Such that, it is impossible to have a public key and create a private key out of it and corrupting the network would basically mean changing each data stored on each node in the network \cite{zhang2019security}.    

\item \textit{Consensus}: The operation of the blockchain frameworks relies on associated consensus algorithms, which is responsible for deciding the group of active nodes on the network. This makes the validation process for a transaction faster and similar to a voting system \cite{leonardos2020presto}.

\item \textit{Accelerated Financial Settlement}: The blockchain transactions are processed much faster in comparison to the traditional banking systems. This technology enables faster transfer of money to foreign workers and overseas travelers. Smart contracts running on the blockchain also help ensure faster settlement of contractual accounts \cite{zhang2020challenges}. 
\end{enumerate}
\subsection{{Big Data}}
% Regardless of techniques, technologies, and applications, big data has been considered in four main angles of volume, variety, velocity, and veracity.

Big data is typically characterized by 4-V features, including volume, velocity, and variety, and veracity\cite{jindal2020unified, sun2018big}. Here, we briefly describe these features of big data.
\begin{enumerate}
	\item \textit{Volume}: Volume simply means the quantity of data, i.e., whether or not a dataset is considered as big data. Regarding big data processing, one usually faces several challenges, which may include the curse of modularity (i.e., not available to store/load the complete data in memory and hard disk), the curse of class imbalance (i.e., there may exist different data distributions), the curse of dimensionality (i.e., the dataset has many features and attributes) \cite{oussous2018big}. Moreover, data non-linearity, variance and bias, and computing availability are also considered as challenges associated with the \textit{volume} feature of big data. 
	
	\item \textit{Variety}: Variety represents various types of data such as video, text, and audio, which are generally composed of structured data, semi-structured data, and unstructured data. The major challenges caused by variety may include data locality, data heterogeneity, dirty and noisy data \cite{jagadish2014big}. Here, data locality expresses that the complete data cannot be stored in a data center and is typically distributed over a large number of physical locations. Data heterogeneity is referred to as various heterogeneous sources of data, thus having different data types, formats, models, and semantics. Dirty and noise data means that the data can contain noise and dirty, which would be caused by data collection methods, data sources, and generation time.  
	
	\item \textit{Velocity}: Velocity refers to the generation speed of data, i.e., how fast the data is generated to meet the demand. A massive number of mobile devices will be 13.1 billion in 2023, from 8.8 billion in 2018, which can generate an enormous amount of traffic \cite{CiscoInternetReport}. Other good examples of the unprecedented growth of data are high-definition videos, video gaming, and streaming platforms (e.g., YouTube and IBM Cloud Video). In some literature, this feature is also considered as \textit{variability}, that is, different applications may have different rates of data flow \cite{mohammadi2018deep}. For example, a vehicular crowdsensing system may generate more data in peak hours due to the participant of a large number of vehicles on the road. 
	
	\item \textit{Veracity}: Veracity refers to the quality aspect since the data can be collected from multiple sources, which may include low-quality and noisy samples. It is reasonable since data can be generated by malfunctioning or uncalibrated IoT devices, untrusted devices, and can be transmitted to the data center via fading and dynamic wireless environments \cite{alsheikh2016mobile}. To improve the quality and analytical accuracy of big data, the challenges of data provenance, uncertainty, dirty and noisy data should be effectively tackled. 
\end{enumerate}
Big data analytics is about extracting useful information and patterns from the dataset, which are then used for different purposes and to create business and social values. In the literature, this is usually considered as the fifth feature of big data, namely \textit{value}. Big data has found applications in many vertical domains such as smart grid, mobile and e-health, transportation and logistics, and wireless and communication networking. Besides great opportunities, we have a number of technological challenges and issues of big data, for example, big data management, data cleansing, imbalanced system capacities, imbalanced data, data analytics, and learning from data \cite{oussous2018big}. For more details, we refer the interested reader to the survey in \cite{oussous2018big} and the references therein.

\subsection{{Motivations of the Blockchain and Big Data Integration}}

Governments and private organizations are investing heavily in big data and blockchain technologies due to their great potential in solving many real-world problems. In modern life, the customers are more inclined to do the transactions online, and expanding amount of data is being generated every day. This exponential rise in the digital data generated creates new opportunities for industries to understand the customer needs, purchasing patterns and trends of the customers. Big data analytics, which uses data mining and statistical models to analyze massive datasets, is playing a major role in helping the industries to gain insights into the purchase patterns of the customers\cite{reddy2020analysis}. However, the tremendous growth in the big data presented its own challenges. Some of the key challenges of big data are security and privacy issues, dirty data, reliability of the data sources, sharing of the data, etc\cite{rawat2019cybersecurity}. These challenges faced by the big data can be addressed by the unique properties of the blockchain like decentralized storage, immutability, transparency, and consensus mechanisms. The motivations of integrating blockchain with big data are discussed as follows.
\begin{itemize}
% \subsubsection{\textcolor{black}{Improving Big Data Security and Privacy}}
\item{\textit{Improving Big Data Security and Privacy}}:
As the number of devices connected to the Internet is growing day by day, the quantity of the data stored at third party locations like cloud is increasing rapidly. \textcolor{black}{This brings new challenges like data breach or threats caused by curious third parties \cite{tang2017big}.} The traditional security solutions like  firewalls cannot address this issue of big data since the organizations have no control over the data as it is not stored within the network perimeter of the organizations. The usage of blockchain to store the big data has the potential to address this issue. The encrypted and decentralized storage of the data in the blockchain network makes it very difficult for any unauthorized access to the data. 

% \subsubsection{\textcolor{black}{Improving Data Integrity}}
\item{\textit{Improving Data Integrity}}:
There exists a likelihood of people tampering the records in big data to influence the prediction of big data analytics in their favor. The immutability property of the blockchain ensures that it is next to impossible to tamper with the data stored in the blockchain network. If someone wants to modify the data in the blockchain network they have to modify the data in at least 50\% of the nodes in the blockchain network, \textcolor{black}{which is nearly impossible in practice.} Also, the immutability property of the blockchain ensures that data stored the blockchain network is reliable. 

% \subsubsection{Fraud Prevention}
\item{\textit{Fraud Prevention}}:
The existing big data solutions rely on the analysis of patterns in the historical data to detect fraudulent transactions. Hence big data cannot solve the problem of fraudulent transactions in the financial sector. The storage of the big data in blockchain enables the financial institutions to monitor each transaction in real time, hence allowing them to assess the potentially fraudulent transactions on the fly. \textcolor{black}{As a result, the integration of blockchain in big data can help the financial institutions to prevent the frauds to protect their customers.}

% \subsubsection{Real-Time Data Analytics}
\item{\textit{Real-Time Data Analytics}}:
Since the blockchain stores every transaction, it makes the real-time analytics of big data achievable. The banks and financial institutes can settle the cross-border transactions including large amounts in near real-time as the blockchain integrated big data analytics enables the financial institutes to settle the transactions quickly. Also, banks can monitor the changes in the data in real time, thus enabling them to make decisions like blocking of the transactions in real time.     

% \subsubsection{Enhancing Data Sharing}
\item{\textit{Enhancement of Data Sharing}}:
The integration of blockchain with big data helps service providers to share the data to other stakeholders with minimal risk of data leakage. Also, if the big data generated from the different sources is stored in blockchain, the repetition of the analysis on the data can be eliminated as each experiment carried out is recorded in the blockchain.  

% \subsubsection{Enhancement of the Quality of Big Data}
\item{\textit{Enhancement of the Quality of Big Data}}:
Data scientists spend most of their time on data integration as different sources follow different formats in data collection. By using blockchain for data storage, the quality of the data can be improved as it is structured and complete. Hence, data scientists can work on the quality data to come up more accurate predictions in real time. 

% \subsubsection{Streamlining the Data Data Access}
\item{\textit{Streamlining the Data Data Access}}:
\textcolor{black}{The use of blockchain would simplify the life cycle of big data analytics} by online streamlining the data access. Indeed, by involving multiple departments in an organization in a common blockchain, authorized users can get access to the secure, trusted data without having to go through several checks.   
\end{itemize}

\section{{Blockchain Services For Big Data}}
\label{Sec:Blockchain_Services_BD}
\begin{figure*}[t]
	\centering
	\includegraphics[width=1.00\linewidth]{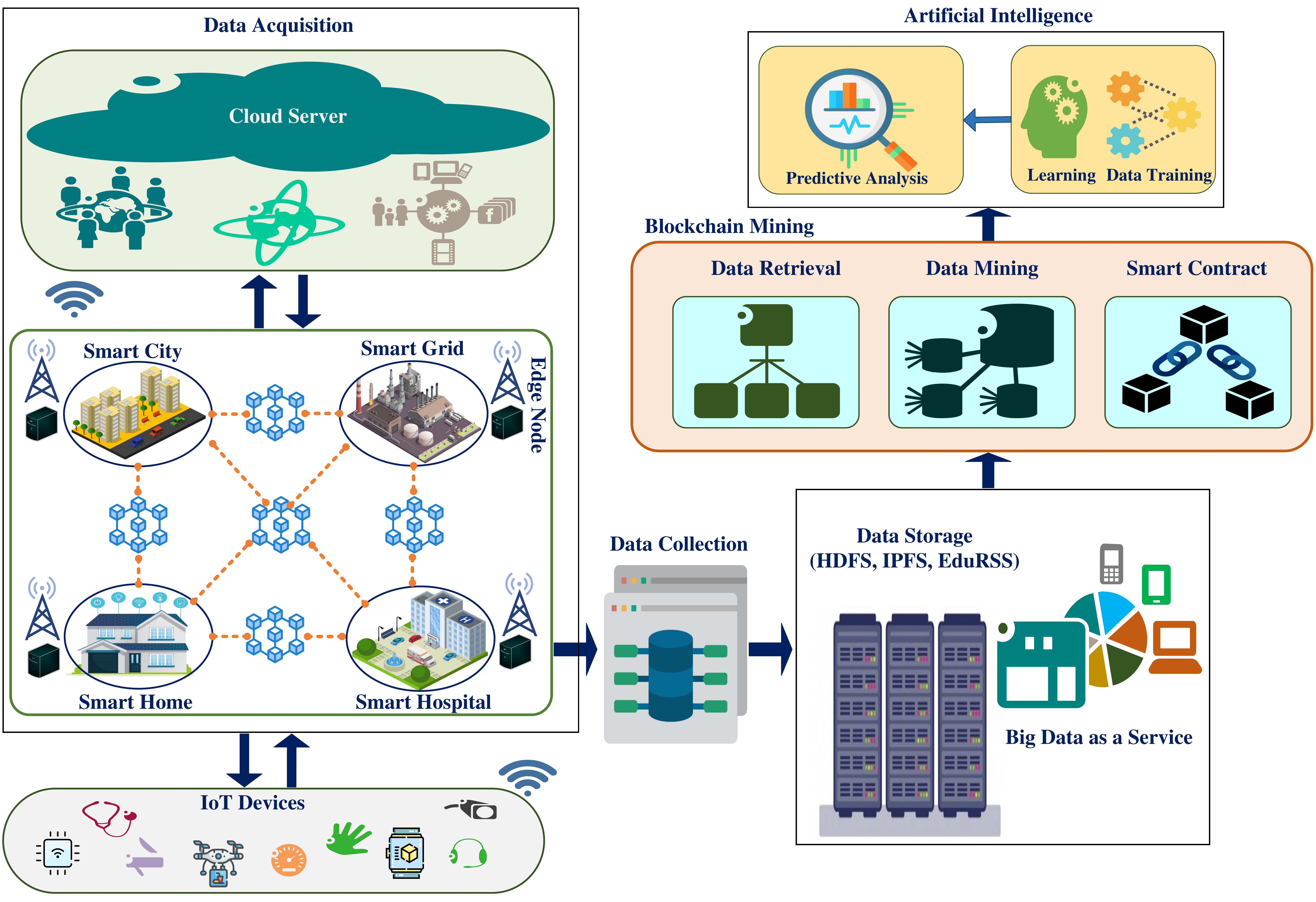}
	\caption{An overview of blockchain services in big data environment.}
	\label{Fig:2}
\end{figure*}

The big data technology has grown tremendously as large corporations and organizations use advanced analytical tools to store, visualize and analyze data. However, due to the enormous data utilization and data transmission, big data security is a major challenge. \textcolor{black}{Cloud computing has been widely used for big data services despite some security concerns.} Some third-party applications and intruders can easily perform malicious activities such as stealing sensitive data, crashing the server when proper security mechanisms are not used \cite{sharma2020blockchain}. Big data faces challenges from a variety of perspectives, such as data collection, data sharing, data storage and data analysis. In this section, we survey the blockchain-based approaches and services for big data. An overview of blockchain services in big data environment \textcolor{black}{such as big data acquisition, big data storage, big data analytics and big data privacy preservation} is depicted in Fig.~\ref{Fig:2}.

\subsection{{Blockchain for Big Data Acquisition }}
In general, big data applications acquire data from diversified sources in a different format (unstructured data). These data cannot be processed in the native form. Therefore, the data must be converted to a structured format from which various predictions on the application domain can be made. Blockchain, with its capability of handling vast data effectively, provides structured data for making predictions. Blockchain ensures data integrity through consensus algorithms thereby mitigating the data attacks. Here, we analyze two subdomains in blockchain services for big data acquisition, including blockchain for secure big data collection and blockchain for secure big data transmission/sharing.

\subsubsection{{Blockchain for Secure Big Data Collection}}
Nowadays, big data applications have gained popularity but faced major security issues and challenges. Data collection is a very important task in the life cycle of data processing. Suspicious sources of data and communication links allow the data collection to expose to various malicious attacks and threats. Therefore, secure data collection methodology is vital for various data applications. Several research works have been done so far to provide secure data collection. \textcolor{black}{For example, a secure big data collection scheme based on blockchain is introduced for mobile crowdsensing (MCS) \cite{bodkhe2020blockchain}.} Due to the rapid growth of portable smart mobile terminal devices such as mobile terminals (MT) and sensors, MCS has been efficiently applied for industrial Internet of thing (IIoT) environment. A MCS framework is developed with cloud servers and a set of MTs. The MCS servers publish some set of tasks related to sensing and choose MTs in the particular area to complete the tasks. The main challenge in performing data collection is limited energy resource in MT, the range of sensing devices and secure data sharing between MTs. A framework was proposed by \cite{liu2018blockchain} to overcome these  challenges using blockchain and deep reinforcement learning (DRL). It provides energy efficient collection of data and security for data sharing in a distributed environment. The distributed blockchain based DRL approach for each MT provides extensive data collection and maximum range for sensing devices. An Ethereum blockchain platform is used to provide data reliability and security while MTs share the data. Ethereum maintains a secure ledger and shares with the cooperating MTs without a trusted third party. The proposed framework provides solutions for various attacks such as majority attack, device failure, eclipse attack, etc \cite{liu2018blockchain}.

\subsubsection{{Blockchain for Secure Big Data Transmission/Sharing}}
\begin{figure}[t]
	\centering
	\includegraphics[width=1.00\linewidth]{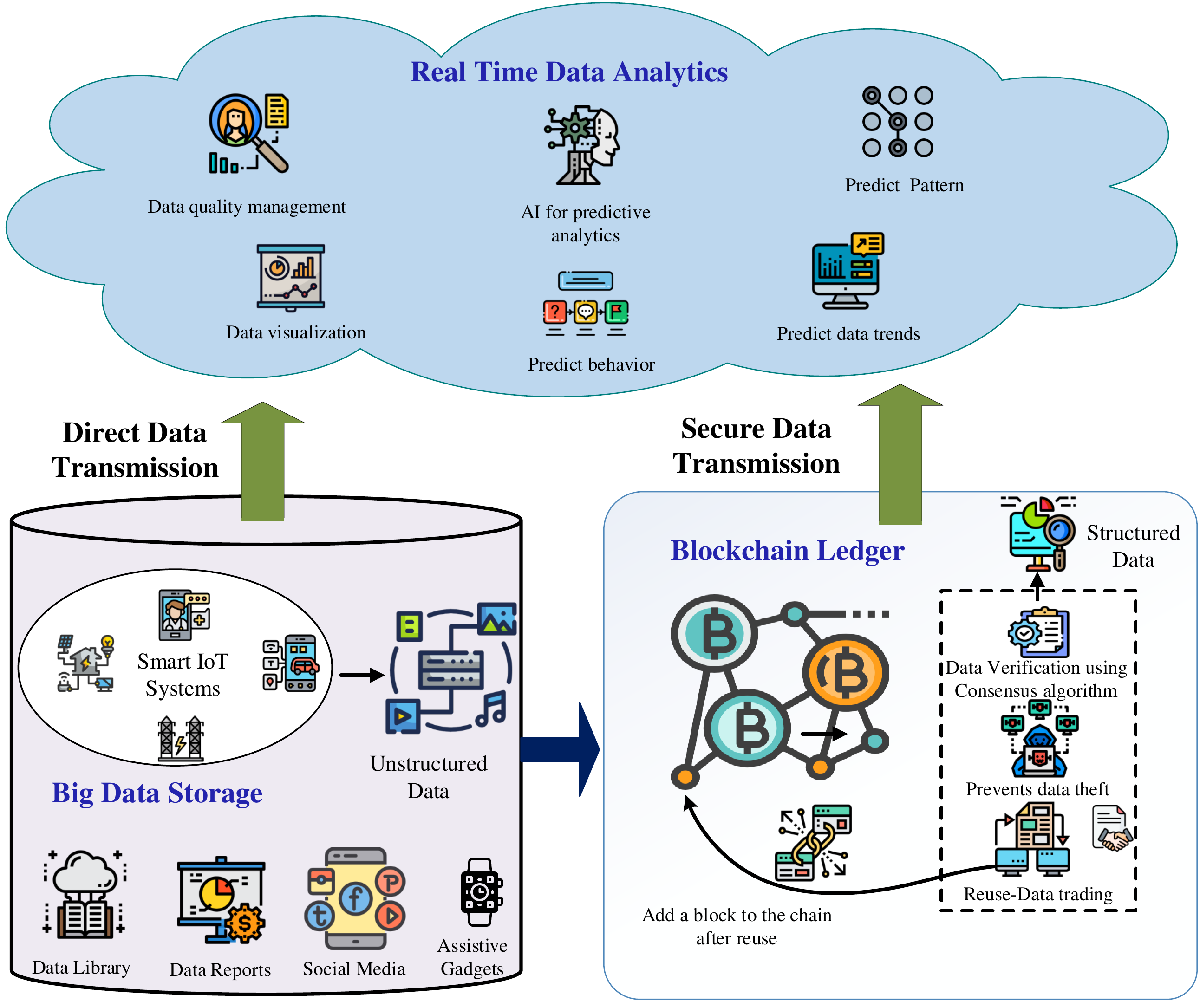}
	\caption{Secured blockchain services for big data processing.}
	\label{Fig:1}
\end{figure}

Blockchain with its decentralized and immutable nature is able to provide secure big data transmissions. It also supports reliable data sharing from data sources to data analytics, aiming to solve security and privacy issues remained in traditional data transmission protocols. Blockchain can ensure big data training and prevent data theft to facilitate big data transmissions. Data can be recorded from ubiquitous sources such as data reports, data libraries, social media, or assistive gadgets. Then, they are added to the blockchain with signature and hash values before sharing with data analytic services in which both data source owners and data analytic users can trace and monitor the data sharing flow over the network which in return provides high transparency and reliable data sharing. An example of big data transmission model with blockchain is illustrated in Fig.~\ref{Fig:1}. 

In the literature, there are some research efforts devoted to use of blockchain for supporting big data transmissions and sharing. 
The emergence of edge computing has seen an increasingly vast amount of data on edge nodes, allowing end-users to optimize latency and the processing time. However, sharing sensitive information without proper authorization is a challenging task. The work in \cite{xu2018making} introduces a blockchain model to share reliable data at the edge node. During this process, the authors pay attention to the reduction of the computational process at the edge nodes using proof-of-collaboration. Besides, to reduce response time and storage overhead, authors introduced a blockchain-based futile transaction filter algorithm that accesses data from the cache layer rather than the storage layer. Finally, the authors proposed express transactions and hollow blocks to increase the efficiency of the network in the proposed model. Express transactions with smart contract is developed in order to support the validation of transactions that are not occurring at the same time period. Also hollow block helps to diminish the redundancy that occur in block generation thereby increases the resource efficacy of the network. The experimental results demonstrate that the proposed model decreases 90\% of computing resources, 95\% of storage resources and 27\% of network resources.
The immense growth of the cyber-physical system helps in providing faster information services and real-time sensing. Big data sends information to the cyber-physical system, which utilizes the radio spectrum. There is incredibly huge competitiveness in the spectrum auction and restricted license-free spectrum access. In \cite{fan2020blockchain}, the authors propose a blockchain-based solution for license-free spectrum access using smart contracts, which facilitates transferring non-real-time data in a secure manner. The proposed framework with edge node aimed to reduce latency provides a blockchain-based protocol, which improves the transaction process in a safe mode where multiple channels are created for spectrum, and each channel is allocated with dedicated blockchain. During the process,  two blocks are created, namely key block and micro block, where key block ensures to select an efficient spectrum license holder, and the micro block takes the responsibility of maintaining all the transaction details. Finally, the valid and authorized node gets the spectrum license from the key block, and the node maintains the license until the key block identifies the next holder.  However, we found that the protocol uses PoS-after-PoW for generating the key block to select a user, which requires high computation cost and consumes more number of resources.

\subsection{{Blockchain for Big Data Storage }}
\subsubsection{{Blockchain for Secure File Systems}}

There are several cloud based services available to store and access files from anywhere on any machine. Users, particularly organizations are hesitant to store sensitive information on the system managed by a third party. Even though encryption of files before storing to the cloud is one of the solutions but still some challenges are faced by the cloud provider in terms of security. At present electronic information system are most popularly used in medical treatment. Volumes of data are produced every day such as medical images, medical records, diagnostic reports, etc. Electronic medical information can affect the treatment of the patients and the experience of the patient is shared with other medical academies. If the shared patient medical data is illegally misused, the privacy of the patient is compromised. Some mechanisms should be adopted to control the access of medical data. Blockchain integrated with interplanetary file system (IPFS) provides the solution for these kinds of security problems. IPFS is a decentralized storage platform developed to address the issue of file redundancy. It defines a unique hash value for the stored file and the user is allowed to find the file based on the hash address. Attribute based encryption method is applied to the medical data before it is stored in the cloud storage. The private key of the user is associated with their attributes and ciphertext with their policy. Any user can perform decryption on the ciphertext if the private key of the user satisfies the access policy available in the ciphertext. Also, blockchain is used to record the data storage and retrieval process. The hash value of medical storage data is stored in the blockchain to provide evidence for the authenticity of user verification. The decentralized blockchain framework  helps to provide security for the file storage and avoids single point of failure \cite{sun2020blockchain}.

\subsubsection{{Blockchain for Secure Database Management}}
Data which is stored in various types of database management system is vulnerable to attacks from internal and external sources. Database tampering detection methods were used to detect the malicious updates in the databases. It uses single-way cryptography hash functions along with digital watermarking to identify data misuse. But the method cannot be applicable to distributed databases. A blockchain based solution is applied to store the data on distributed databases and detect the malicious user transactions. Blockchain avoids data tampering by applying time stamping. Virtual shared ledger is incorporated to store the history of transactions. All the transactions are recorded in block and each block is interconnected with each other with cryptographic hash values. When the data available in a block is updated by a malicious attacker, the hash value of the block gets updated and the block becomes invalid. A blockchain based scheme, namely education records secure storage and sharing scheme, is proposed for privacy preserving and secure storage of education records. The scheme integrates storage servers, cryptographic algorithms and blockchain to develop a safe and reliable environment. Blockchain is integrated to provide reliability and security to the education database. The smart contracts in blockchain are applied to control the data sharing and storage process. The educational records are linked with the hash information stored in the blockchain to provide security to the stored data. Cryptographic algorithms and digital signatures are used to maintain the encryption of the records \cite{li2019edurss}.

\subsubsection{{Blockchain for Big Data Storage Infrastructure}} In the past few years, big data has grown into a new standard which provides huge amount of 
data and prospects to enhance the decision making applications in science and engineering. At the same time, it faces challenges in storing, processing and transmitting the data. Cloud computing offers basic support to address the issues with shared resources such as networking, storage and computing. The increased readiness of data in AI provide opportunities in the healthcare industry.  The present machine learning (ML) algorithms transform a person data to medical data for data analytics preventing the patients access to their medical data. Blockchain can provide security solutions to motivate the biomedical research and allow the patients to access and control their personal data along with the capability of monitoring their health records. Blockchain provides decentralization facility for a transparent and secure distributed personal data \cite{yang2017big}.

\subsection{{Blockchain for Big Data Analytics}}
\subsubsection{{Blockchain for Secure Data Training}}
The development of edge and cloud computing has increased the amount of data in various scenarios. Several ML and deep learning (DL) methods are applied for effective data analysis. Support vector machine (SVM) is one of the popular ML methods applied for its efficiency and accuracy. In vehicular social networks, data are gathered from various entities namely social network companies, vehicular manufacturers and vehicle management agencies. Data from various data sources normally differ in the attributes. When training with SVM classifier, the entities face problem of the data with inadequate attributes due to the diversity of sources. Therefore, various entities must share data to integrate the dataset with multiple attributes and train the classifier. Data privacy issue occurs due to sharing of data from various entities. A privacy preserving blockchain based SVM training method was proposed for vertically partitioned dataset from various data providers. In this method, a blockchain consortium and homomorphic cryptosystem were developed to implement a secure training platform without the need of a trusted third party. The training operations are performed over the original data locally and the interactions between the entities are secured by the homomorphic cryptosystem and blockchain consortium. Blockchain consortium helps to build a public and secure data sharing environment for effective communication between the entities when they share the attribute values \cite{shen2019secure}.

\subsubsection{{Blockchain for Secure Data Learning in AI Algorithms}}
The extensive construction and generation of data from sensors, social media, web and IoT devices resulted in the growth of Artificial Intelligence (AI) techniques. The data can be applied to ML and DL algorithms for the purpose of data analytics. These methods depend on centralized server for training purpose and it leads to tampering of data. Thus the decisions obtained from AI are erroneous and risky. Therefore the decentralized AI came into existence to solve this problem and it is integration of blockchain and AI. Several limitations of blockchain and AI are solved by integrating these two technologies. AI techniques depend on data to learn, gather and provide decisions. These techniques perform better when the input data are gathered from various secure, reliable and trusted data repositories. Blockchain provides secure environments through distributed ledger in which data can be recorded and transacted \cite{chowdhury2019comparative}. Here, data are stored with high resiliency and integrity in blockchain and cannot be tampered. When smart contracts are used for learning purpose in AI algorithms to obtain decisions and analytics, the results can be undisputed and trusted. Therefore, the integration of blockchain with AI can provide immutable, decentralized and secure environment for learning the highly sensitive data. This integrated framework provides substantial development in various domains such as banking, medical, financial, personal and trading \cite{salah2019blockchain}. 

\subsection{{Blockchain for Big Data Privacy Preservation}}
\subsubsection{{Blockchain for Privacy Preservation in Big Data Processing}}
Due to the rapid increase in generation of data, privacy preserving has become a main concern nowadays. In this era of big data, the data is regularly being gathered and examined which leads to commercial and innovation growth. Big organizations and companies utilize the collected data to provide better customer services, optimize the decision process and forecast the future developments. Thus data has become a valuable asset in recent days. Big data is widely applied in smart city environment for extensive monitoring of city traffic and maintenance, ensuring quality of air and water etc. A blockchain based model is proposed \cite{hirtan2018blockchain} for privacy preserving in intelligent transportation system (ITS) for in-car navigation system in smart city environment. The model applies offline blockchain based storage in which all the sensitive information from the users are stored in a secure manner. The sensitive data is encrypted with the help of shared key associated with a group of cars. Users can use various security features such as sharing details about speed enable and disable, location enable and disable, etc  \cite{hirtan2018blockchain, sharma2018blockchain}.
\subsubsection{{Blockchain for Privacy Preservation in Big Data Storage}}
The big data era is threatening the user privacy in various digital scenarios. Third party organizations are benefited in the management of user data by gathering, analysing and managing the huge amount of user personal information. These services provided by the third parties are prone to security breaches and data misuse without the knowledge of the users. Blockchain provides various solutions to the challenges faced by the user data. User transactions in blockchain do not face privacy concerns and users are provided with options to control their personal information. The details about when, by whom, which and what personal information is revealed in each transaction. Privacy preserving solutions are emerging for blockchain built on crypto-privacy methods to allow the users to become unidentified and gain control over their personal information during their digital transaction in ledger \cite{bernabe2019privacy}. The various services provided by cloud environment for big data, challenges and blockchain based solutions are tabulated in Table \ref{tab:sec3}.
\begin{table*}[h!]
\caption{Services provided by cloud environment for big data, challenges and  blockchain based solutions.}
\label{tab:sec3}
\centering
\resizebox{\textwidth}{!}{%
\begin{tabular}{|l|p{3.8cm}|p{4.9cm}|p{7cm}|}
\hline
\multicolumn{1}{|c|}{\textbf{Ref}} &
  \multicolumn{1}{c|}{\textbf{Cloud based services}} &
  \multicolumn{1}{c|}{\textbf{Challenges faced by Big data}} &
  \multicolumn{1}{c|}{\textbf{Solutions   provided by Blockchain}} \\ \hline
  \cite{liu2018blockchain} &
  Data collection. &
  Data collection is exposed to various malicious attacks and threats. &
  Blockchain provides energy efficient data collection and secure data sharing environment using Ethereum. \\ \hline
  \cite{xu2018making} &
  Data transmission/sharing. &
  Lack of authorization for data sharing in edge nodes and response time is more. &
  Blockchain based futile transaction filter algorithm helps to access data from cache layer instead of storage layer and helps to reduce response time and storage overhead. Smart contracts are used for authorization. \\ \hline
  \cite{sun2020blockchain} &
  File storage system. &
  Unauthorized access to the electronic file system. Privacy, security and redundancy problems. &
  Blockchain integrated with IPFS provides the solution by implementing decentralized platforms to solve file redundancy problems and provides security to the file storage system. Hash value of data is stored in blockchain to provide authenticity to the users and  an attribute based encryption method is applied   before data storage in cloud. \\ \hline
  \cite{li2019edurss} &
  Database management system. &
  Data stored in distributed database is exposed to internal and  external attacks. &
  Blockchain overcomes data tampering using time stamping method. Virtual shared ledger is applied to store the transaction history. Database transactions are recorded in block and each block is interconnected with each other using cryptographic hash value. Blockchain based solution integrates storage servers, cryptographic algorithms for a reliable database access. \\ \hline
  \cite{shen2019secure} &
  Data training/learning process. &
  Various entities share data to integrate the dataset with various attributes and train the ML classifier. Data privacy issue occurs while sharing data from various entities. &
  Blockchain consortium and homomorphic cryptosystem provide a secure training platform without the intervention of a trusted third party. Blockchain provides a secure environment for communication between the entities. \\ \hline
  \cite{ bernabe2019privacy} &
  Data privacy preservation. &
  User privacy is an issue in digital scenarios in big data era. Services provided by third parties are exposed to security breaches and data misuse. &
  Blockchain provides immutable, verifiable and decentralized ledger to record the transactions in digital scenarios.It provides facilities to the user to control their personal data. Crypto-privacy methods are applied to solve privacy preserving problems. \\ \hline
\end{tabular}%
}
\end{table*}

\section{{Blockchain Big Data Applications and Projects }}
\label{Sec:BlockchainBD_Applications}
Blockchain technologies have gained immense momentum with its varied applications in various spheres of life. The technology is still going through its phase of infancy and is being experimented for providing solutions to  various challenges pertinent to security, data ownership, decision support systems, identity verification and decentralization. Our present generation is traversing through an era of overwhelming volume of digital data, being generated by man and machines. Hence there emerges a desperate need to store, organize, process and analyze this big data where the use of blockchain technologies has a potentially significant role to play \cite{kim2020advanced}. As an example, maintaining data ownership, data transparency and management of access control has always been a major challenge. Blockchain technology resolves this issue by storing access policies to personal data in the blockchain framework. By using the blockchain technology, a decentralized personal data management system is created by implementing a protocol allowing users to own and manage their data. The dependency on third party is completely eliminated allowing organizations to focus more on data utilization rather than security management and compartmentalization \cite{chen2019design}. The application of blockchain in combination with big data is visible in two segments - data management and data analytics. The various blockchain based big data applications are summarized in Table \ref{table_app}. In case of data management, blockchain technologies being sure and distributed, are implemented to store important data. It can also evaluate data authenticity and stop tampering of sensitive data. In the applications of data analytics, blockchain is used to analyze trading trends, prediction of potential customers, diseases or business partners \cite{monrat2019survey}.

\subsection{{Blockchain Big Data in Smart City}}
The rapid urbanization have led to the development of smart cities which requires efficient and intelligent solutions for its transportation, administration, environment and energy optimization. The integration of IoT, big data and energy efficient Internet technologies has the capability to provide such infrastructural solutions required for the smart city life. But there are numerous problems related to inferior security, reliability, maintenance, adaptability and costs. The blockchain technology caters to such needs having transparency, energy efficiency, space, recover-ability and maintenance of the IoT devices. The study in \cite{ali2018applications} discusses the use of hash, asymmetric encryption, consensus algorithm, a blockchain structure and a Merkle tree in ensuring a tamper free transaction. This framework has blocks interlocked with one another within the block itself with the help of a Merkle tree which makes it even more secured for performing seamless transactions. 
The recent years have also witnessed a surge in the development of big data based auditing systems termed as third party auditors (TPAs). The TPAs are centralized frameworks which are subjected to security issues within the cloud environment. Blockchain technologies have been used to create decentralized TPAs for smart cities with enhanced security and reliability. This framework is named as Data Auditing Blockchain (DAB), the entire audit history is traced and also allows owners to audit their files at any point in time. It also includes the feature of batch verification of various auditing proofs ensuring security and prevention of privacy \cite{yu2018decentralized}.    
In \cite{rahman2019blockchain}, a blockchain based infrastructure is presented that provides secured spatio-temporal smart contract services. The framework provides sustainable IoT based shared economy in smart mega cities. The huge generation of big data has created the need to collect, analyze and utilize the same for autonomously predicting any risky or exceptional events from occurring. The framework consists of device-to-device (D2D) communication systems and fog nodes installed onsite to enable the blockchain and other offline operations. A three-tier architecture is used for supporting shared economy services in the blockchain based smart city environment. The client tier includes the smart applications, IoT and associated infrastructures. The client tier communicates with the mobile edge tower through WiFi, ZigBee, 5G and other related technologies. The MEC tower hosts the blockchain nodes, data storage client, related databases and cloudlet applications, thereby manages the load efficiently. The data from the blockchain, IoT and social network are finally fed into the AI engine for performing sophisticated analysis such as digital forensics, emotion extraction and various others. 

\subsection{{Blockchain Big Data in Smart Healthcare}}
\textcolor{black}{Recent advances in the healthcare sector have led to a drastic rise in medical data generation.} These data are extremely important for diagnosis, predictions and treatment purposes. Healthcare professionals have recently started focusing on the use of IoT and related wearable technologies wherein sensors, devices are vehicles are connected through the Internet providing services for the benefit of mankind. As an example the remote patient monitoring system is a common device for treating elderly patients in particular. Although these technologies have enormous benefits but have aforementioned security issues while transferring and logging of data transaction information. But these issues have possibilities of extreme violation of data security and privacy. The use of blockchain is a potential solution that would provide security and efficiency in analysing data but it is costly and lags energy optimization. The study in \cite{dwivedi2019decentralized} proposes a framework that resolves such issues using public key, private key, light weight cryptographic techniques in integration with blockchain technology. The framework thus provides an access control of medical records for patients with improved privacy and security. 
In \cite{mcghin2019blockchain} a secured smart health care system is proposed  using blockchain. The various private data, public data and related sensitive information are captured using sensors and then encrypted using blockchain technologies. These types of information are further stored in a distributed format rather than centralized cloud storage systems, which can be accessed only by authorized individuals having approvals from patients. Similarly, the healthcare professionals seeking to access the patient records need to send request to the patient and once real time notification is processed, information is available to them. \textcolor{black}{All the entities such as IoT devices,} Electronic Health Records (EHRs), Encryption/decryption system, blockchain mechanisms in this framework remain connected through wireless sensor networks (WSN) to conduct seamless yet secured communication. 
In \cite{vyas2020integrating}, a private blockchain framework is proposed using Ethereum protocol wherein the sensors communicate with the smart devices. These smart devices call smart contracts which keep records of all events on the blockchain. Thus, these smart contract systems help in monitoring patients in real-time and also send notifications to healthcare professional when medical interventions are required. The saved records are secured, due to the connectivity in the blockchain which provides authentication and eliminates possibilities of data tampering of EHRs.

\subsection{{Blockchain Big Data in Smart Transportation}}
Transportation helps to move human beings and goods from one location to another. Although the application of blockchain has the immense potential towards benefiting the transportation sector, but individuals in this sector are not well informed about this emerging technology. Various other technologies namely Mobility as a Service (MaaS), IoT, AI and DL have converged with blockchain technologies to revamp the traditional approach involved in transportation. The automotive sector has also used blockchain technologies for developing intelligent transportation systems and offer services like remote software based vehicle operation system, automated insurance services, smart charging and cab sharing services \cite{wang2019blockchain}. 

Blockchain technologies have seen a rapid growth due to its potential to revolutionize intelligent transportation systems (ITS). Such developments can be used to create secured, reliable and autonomous ITS ecosystems with optimized usage of relevant infrastructure and resources. As an example, the study by \cite{astarita2020review} presented a seven layer conceptual model for ITS that would help in characterizing the architecture and major components in a blockchain based system.   The physical layer holds the different vehicles, devices and assets relevant to ITS. The main aspect of this layer includes the use of IoT for providing enhanced security and privacy for the blockchain based transportation systems. The data layer provides the data blocks and associated encryption algorithms, hashing algorithms and Merkle trees. The network layer defines the process involved in distributed networking, data forwarding and authentication. The packaging of the consensus algorithms is done by the consensus layer followed by the incentive layer which specifies the mechanisms for issuance and allocation of coins to nodes in the blockchain network. The contract layer constitutes algorithms and smart contracts that activates the process of data storage in the blockchain. Finally, the application layer encompasses the scenarios and use cases of blockchain based ITS. 
Security is often a major concern in vehicle communication systems. The study in \cite{lei2017blockchain} presented a secure key management framework for accomplishing network security. The study utilizes the role of security managers who capture vehicle departure data and encapsulate the blocks to transport keys and later implement rekeying to the vehicles within the secured domain. The framework proposes an efficient key management system for key transfers among the security managers in a heterogeneous vehicle communication network architecture.
In addition to security provision, blockchain technologies play a significant role in privacy protection of ITS especially in car navigation. The framework proposed in \cite{hirctan2020blockchain} is based on an offline blockchain storage system wherein all sensitive data extracted from the users are stored and later shared using specific encryption keys relevant to a particular car cluster. The system uses two major applications namely the client application installed in the users smart phone and the main application installed at the server side. It is assumed that the smart phone and server are configured securely and security policies both simple and complex are used depending on user types for data sharing. All clients in the network are grouped into clusters depending on the location to optimize the use of computational resources, reduce network delays and overheads. The system allows users to define the privacy policies and they are later automatically implemented at the client application that provides accurate transportation routes. 
Blockchain technologies can potentially solve various problems relevant to car insurance. The insurers with the help of this technology will be able to track their claims seamlessly by searching the trusted ledger. The study by \cite{li2018blockchain} presented a prototype framework for fine-grained transportation insurance services where the premium was calculated based on vehicle usage and behavior of the driver. These information were collected by streaming IoT data collected using mobile sensors. This unique framework initiated transparent insurance and also motivated drivers to drive safely in-order to achieve insurance incentives. The mobile GPS sensors in this framework were strategically placed in vehicles for continuous monitoring of their GPS location. The GPS trajectory data were further uploaded to the public cloud or data center using the IoT suite and later saved in the GIS database. The IoT messages trigger spatio-temporal data analytic function to extract driver behavior and vehicle usage data. These data get saved in the distributed ledger system on the blockchain ensuring transparency, trace-ability and safety. In case of Ethereum based framework the premium evaluation is done based on driver behavior and vehicle usage which is tokenized according to varying risk levels through a fine grained process. 

\subsection{{Blockchain Big Data in Smart Grid}}
Blockchain technologies can contribute significantly to improve the efficiency of practices and processes in the energy sector. Blockchain integrated with big data has the ability to accelerate the speed of development of IoT platforms and digital applications thereby innovating the P2P energy trading and decentralization services. The present energy systems are experiencing radical transformations due to the advancement of distributed energy resources and use of information and communication technologies. The blockchain architectures have the capability to solve issues relevant to controlling and managing of decentralized energy systems and micro-grids \cite{andoni2019blockchain}.
Smart grid is a technology that makes electrical power grids more efficient, robust and less pollutant. The advanced metering infrastructure (AMI) is one of the major components in smart grid architecture that ensures two way communication between users and the utility device, by installing a smart meter at the user end. Key management plays a major role in this process and most of the traditional architectures depend on a single entity to distribute the keys and maintenance. The study by \cite{baza2020blockchain} proposes a distributed key management system to maintain optimum security in the smart grid system. A key agreement protocol is proposed between the utility and the smart meter followed by the use of a distributed multi-case key management scheme which allows group members to effectively manage their group communication. The blockchain architecture enables distributed entities to interact with each other in the distributed P2P network ensuring security, scalability and efficiency. In \cite{fan2019consortium}, a data integration and regulation system is proposed based on consortium blockchain. A signcryption algorithm is implemented to multidimensional data acquisition and the receivers in the blockchain framework. As part of the regulation process, the control center, the grid operator and the grid supplier receive fixed blocks from the blockchain and later obtain plaintext from the decryption process. At the outset, multidimensional data are analyzed by the relevant receivers. This results in creation of control pieces. These control pieces takes care of the security and data integrity aspect thereby reducing communication costs. In \cite{jindal2019guardian}, a blockchain based demand response management system is proposed. This system is termed as GUARDIAN and is capable of taking trading decisions pertaining to the energy sector. The system is extremely secured and also contributes significantly towards load management in the residential, industrial and commercial sector. The minor nodes in this framework termed as block verifiers, are selected using their specific power consumption and processing power capability. These nodes help in the authentication of energy transactions in the smart grid network. The transaction process in the proposed framework is initiated by the end user which creates the block of transaction for energy trading. The miner nodes validate these blocks, add them to the blockchain and they become eligible to be part of the energy trading. This helps in achieving security and eliminates unauthorized entries in energy trading. 

\subsection{{Blockchain Big Data Projects}}
\begin{figure*}[t]
	\centering
	\includegraphics[width=1\linewidth]{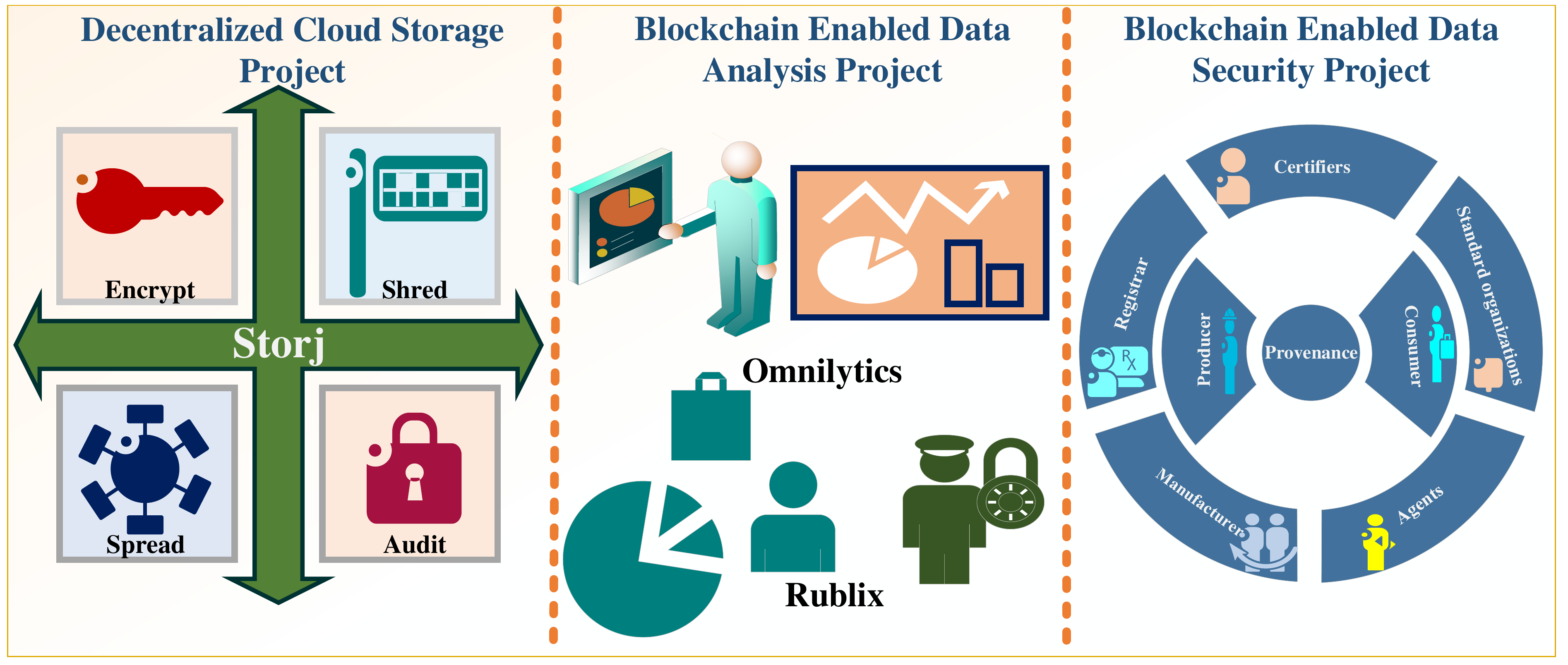}
	\caption{Popular blockchain big data projects.}
	\label{Fig:3}
\end{figure*}

Blockchain is a technology that empowers cryptocurrencies such as bitcoin and ethereum. On the other hand, big data is an advanced concept of data science which involves larger dataset with great variety, size and velocity. These datasets are analyzed to reveal interesting patterns, association and trends. Interestingly, blockchain is a type of distributed ledger that records transactions in a way that cannot be altered. There is an immense trust factor associated with blockchain that eliminates the need of third parties to regulate transactions ensuring the data is immutable. Blockchains have many applications in data science where data integrity is maintained while performing data analysis and data sharing \cite{lao2020survey}. The benefits of the application of blockchain are applied in three areas namely for decentralized data storage, performing blockchain enabled data analysis and finally in maintaining blockchain enabled data security as shown in Fig. \ref{Fig:3}. Some of the blockchain projects for big data applications are discussed below.

\subsubsection{{Storj}}
Storj is an end-to-end decentralized storage project which utilizes the excess hardware and bandwidth capacity, enabling peer to peer authentication of storage contracts between the providers and the users \cite{zhang2019frameup}. The process involves encryption of the files at the client side which are the split into pieces termed as "shards". These shards are later stored three times to maintain backups at the farmer side. The client only has access to the data which provides additional security than the traditional centralized cloud services. The Storj cryptocurrency allows renters to check on the farmers files and also pay for the maintenance of this storage system. The renters pay only for the space used without any additional fees pertaining to user requirements of setup costs. 
\subsubsection{{Omnilytics}}
Omnilytics is a blockchain platform for big data analytics that provides insights for sales, marketing and merchandising industry \cite{moreno2019blockbd}. It uses blockchain, big data analytics, ML, AI and various other technologies to integrate data from different industries. The platform provides data analytics and related services for competitor benchmarking, trend analysis and pricing analysis for the clients. Blockchain is used to empower smart contracts, distributed data finger printing, data exchange and other services to track the trend of data, provide incentives through micropayments. 

\subsubsection{{Rubix}}
Rubix blockchain \cite{kokina2017blockchain} uses the concept of decentralization to integrate the cryptocurrency traders in a common trading platform to authenticate their credibility and predictions. The protocol is based on the transparency and immutability attribute of blockchain in combination with investment data analytics to generate more accurate trading predictions. The traders as a result are ranked based on the accuracy of their predictions wherein the blockchain verifies the traders and incentivizes them based on the content quality.  

\subsubsection{{Provenance}}
Provenance is a blockchain platform mainly used in supply chain management that helps to gather important product information and shares the same in a trusted, secure and accessible manner \cite{kim2018toward}. The blockchain architecture used encompasses of six participants namely - the producer, the manufacturer, registrar, standard organizations, agents like certifiers or auditors and finally the customers. The protocol provides access to information to its consumers on origin of the products, its journey along various points in the supply chain, product quality and its impact on the environment. 

\subsubsection{{FileCoin}}
FileCoin intends to create a decentralized storage network that would allow traders to buy and sell storage in an open market. The FileCoin allows users to rent storage on devices having excess storage spaces using the filecoin cryptocurrency. The clients spend cryptocurrencies for sharing or retrieving of the data and miners earn the filecoins through storage and services of data. When the miners mine a particular block, they need to submit a proof-of-space-time (PoST) to the network, which validates if a storage provider is performing the required responsibilities for storing outsourced data for the stipulated time frame. The filecoin consists of blockchain, retrieval nodes, storage nodes and a native filecoin token. The storage nodes store sealed copies of data and the transactions are recorded by the blockchain. The retrieval nodes fetches and delivers the files to the users abiding to the PoST \cite{L3}. 

\subsubsection{{Datum}}
Datum (DAT) is a decentralized, distributed, high performance and NOSQL platform supported by Ethereum, Bigchain DB and IPFS. It basically enables users to store data anonymously and securely from social network, IoT devices and wearable technologies. The platform also acts as a marketplace for sharing and selling of data by providing Datum users with a unique Datum ID which are managed by the Datum mobile application available for Android and iOS services \cite{L2}. 
The summary of big data blockchain applications is described in Table \ref{table_app}. 

\begin{table*}[h!]
\centering
\caption{Summary of blockchain big data applications.}
\label{table_app}
\resizebox{\textwidth}{!}{%
\begin{tabular}{|c|c|p{4.85cm}|p{4.85 cm}|p{4.8cm}|}
\hline
\multicolumn{1}{|c|}{\textbf{Ref.}} &
  \textbf{Application} &
  \multicolumn{1}{c|}{\textbf{Description}} &
  \multicolumn{1}{c|}{\textbf{Benefits}} &
  \multicolumn{1}{c|}{\textbf{Research Challenges}} \\ \hline
\cite{ ali2018applications} &
  \multirow{6}{*}{Smart City} &
  \begin{tabular}[c]{@{}l@{}}Use of hash, asymmetric encryption, \\ consensus algorithm, blockchain and \\ merkle tree.\end{tabular} &
  \begin{tabular}[c]{@{}l@{}}-Tamper free transactions in IoT devices. \\ -Development of  decentralized, secure \\ and auditable environment for IoT \\ devices.\end{tabular} &
  \begin{tabular}[c]{@{}l@{}}-Maintaining balance between privacy and\\ accountability.\\ -Implement  Blockchain for crowd sensing.\end{tabular} \\ \cline{1-1} \cline{3-5} 
\cite{ yu2018decentralized} &
   &
  \begin{tabular}[c]{@{}l@{}}Secured big data auditing scheme using \\ DAB.\end{tabular} &
  \begin{tabular}[c]{@{}l@{}}-Elimination of  centralised third party \\ auditors.\\ -Improvement in reliability and stability.\end{tabular} &
  NA. \\ \cline{1-1} \cline{3-5} 
\cite{ rahman2019blockchain} &
   &
  \begin{tabular}[c]{@{}l@{}}Use of fog nodes and D2D to enable \\ blockchain.\end{tabular} &
  \begin{tabular}[c]{@{}l@{}}Prediction of   risk or any exceptional \\ event in smart contract.\end{tabular} &
  \begin{tabular}[c]{@{}l@{}}Large scale testing in shared economy \\ scenarios could be implemented.\end{tabular} \\ \hline
\cite{ dwivedi2019decentralized} &
  \multirow{8}{*}{\begin{tabular}[c]{@{}c@{}}Smart \\ Healthcare\end{tabular}} &
  \begin{tabular}[c]{@{}l@{}}Access control of patient records using \\ cryptographic techniques integrated with \\ blockchain.\end{tabular} &
  \begin{tabular}[c]{@{}l@{}}Provides security and privacy for IoT \\ based health monitoring systems.\end{tabular} &
  \begin{tabular}[c]{@{}l@{}}-Resource constraints of IoT acts as a \\ major challenge.\\ -Commercialization in collaboration with \\ industry partners.\end{tabular} \\ \cline{1-1} \cline{3-5} 
\cite{ mcghin2019blockchain} &
   &
  \begin{tabular}[c]{@{}l@{}}-Sensitive information collected using \\ sensors are encrypted and distributed \\ using blockchain. \\ -Implemented in IoT based EHRs systems \\ connected through WSN.\end{tabular} &
  \begin{tabular}[c]{@{}l@{}}Ensures identity verification and fraud \\ detection.\end{tabular} &
  \begin{tabular}[c]{@{}l@{}}Implementation on large scale healthcare \\ data.\end{tabular} \\ \cline{1-1} \cline{3-5} 
\cite{vyas2020integrating} &
   &
  \begin{tabular}[c]{@{}l@{}}Implements Ethereum smart contract \\ where sensors communicate with the \\ smart devices.\end{tabular} &
  \begin{tabular}[c]{@{}l@{}}-Monitors patients in real-time. \\ -Sends alerts for medical interventions.\end{tabular} &
  \begin{tabular}[c]{@{}l@{}}-Large scale adaptation.\\ -Resource utilization.\end{tabular} \\ \hline
\cite{wang2019blockchain} &
  \multirow{10}{*}{\begin{tabular}[c]{@{}c@{}}Smart \\ Transportation\end{tabular}} &
  \begin{tabular}[c]{@{}l@{}}Implementation of Smart contract, track \\ \& trace, fast payment and supply chain \\ finance using blockchain.\end{tabular} &
  \begin{tabular}[c]{@{}l@{}}Data authentication, decentralization to \\ provide knowledge in shipping, logistics,\\ transportation .\end{tabular} &
  \begin{tabular}[c]{@{}l@{}}-Varied successful applications considering \\ transportation engineering still not \\ predominant.\end{tabular} \\ \cline{1-1} \cline{3-5} 
\cite{lei2017blockchain} &
   &
  \begin{tabular}[c]{@{}l@{}}Implements a key management system \\ for key transport in VC systems.\end{tabular} &
  \begin{tabular}[c]{@{}l@{}}Captured vehicle departure data, ensures \\ secured key transport and re-keying of \\ vehicles.\end{tabular} &
  \begin{tabular}[c]{@{}l@{}}-Maintaining of balance between security \\ and privacy.\\ -Pseudonym management can be included.\end{tabular} \\ \cline{1-1} \cline{3-5} 
\cite{hirctan2020blockchain} &
   &
  \begin{tabular}[c]{@{}l@{}}Enables users to travel between locations, \\ make insurance and finance decisions on  \\ blockchain based disruptive technology.\end{tabular} &
  \begin{tabular}[c]{@{}l@{}}Defines privacy policies and resolves \\ issues to transportation route finding, \\ car insurance and tracking of claims.\end{tabular} &
  \begin{tabular}[c]{@{}l@{}}Use of GPS positioning of the users not \\ included.\end{tabular} \\ \cline{1-1} \cline{3-5} 
\cite{li2018blockchain} &
   &
  \begin{tabular}[c]{@{}l@{}}Implements a fine grained transportation \\ insurance service based on vehicle usage, \\ driver behavior using hyperledger and\\ cryptocurrency.\end{tabular} &
  \begin{tabular}[c]{@{}l@{}}Promotes safe driving and unbiased \\ insurance claims.\end{tabular} &
  \begin{tabular}[c]{@{}l@{}}Implementation in large scale in cities\\ and other similar applications.\end{tabular} \\ \hline
\cite{andoni2019blockchain} &
  \multirow{10}{*}{Smart Grid} &
  \begin{tabular}[c]{@{}l@{}}Implementation of blockchain, distributed \\ consensus algorithms in energy industry.\end{tabular} &
  \begin{tabular}[c]{@{}l@{}}Innovations in P2P energy trading and\\ decentralized energy generation.\end{tabular} &
  \begin{tabular}[c]{@{}l@{}}Achieving market penetration and \\ commercial viability.\end{tabular} \\ \cline{1-1} \cline{3-5} 
\cite{baza2020blockchain} &
   &
  \begin{tabular}[c]{@{}l@{}}Distributed key management for AMI in \\ Smart grids.\end{tabular} &
  \begin{tabular}[c]{@{}l@{}}Use of  multi-case key management \\ scheme for security of smart grids.\end{tabular} &
  \begin{tabular}[c]{@{}l@{}}-Scalability to process transaction.\\ -Difficulty in prediction of price due to \\ volatility in supply and demand.\end{tabular} \\ \cline{1-1} \cline{3-5} 
\cite{fan2019consortium} &
   &
  \begin{tabular}[c]{@{}l@{}}Implementation of signcryption algorithm \\ for data integration and regulation system \\ inblockchain framework.\end{tabular} &
  \begin{tabular}[c]{@{}l@{}}-Security, of multidimensional data.  \\ -Reduction of communication costs.\end{tabular} &
  Real-time analysis could be included. \\ \cline{1-1} \cline{3-5} 
\cite{jindal2019guardian} &
   &
  \begin{tabular}[c]{@{}l@{}}Implements GUARDIAN, blockchain \\ secured demand response management.\end{tabular} &
  \begin{tabular}[c]{@{}l@{}}-Enables accelerated decision making \\ for energy trading.\\ -Load management in residential, \\ industrial \& commercial sector.\end{tabular} &
  \begin{tabular}[c]{@{}l@{}}-Testing and deployment of the scheme \\ on larger dataset.    \\ -Optimization  of the scheme to reduce \\ latency and increase network throughput.\end{tabular} \\ \hline
\end{tabular}%
}
\end{table*}

\section{{Research Challenges and Future Directions}}
\label{Sec:Challenges_Future-Directions}
The robust technologies blockchain and big data are evolving in almost all domains. When these powerful technologies are integrated, the integration opens up new research opportunities due to massive data accumulations in today's data centers. Big data is evolving in business organizations today gaining higher profits. Similarly, data held up in the blockchains worth more through its sensitive nature.  Blockchain validates the data ensuring quality in data management, whereas big data analytics makes better predictions on a large quantity of data. These technologies have specific challenging issues to be addressed when used individually and in combination with their adoption. The most prominent challenges are massive data silos in the big data environment that should be secured, ensuring integrity and repudiation in data transactions. Therefore, blockchain with its decentralized framework and secured immutable nature will be an optimal choice.  Indeed, blockchain possesses some challenges to be addressed on its deployment. The goal of integration is to store the massive data on the decentralized ledgers instead of centralized servers with authorized data access \cite{chen2019fade} and allowing the users to share their unused storage on the exchange of cryptocurrencies like bitcoins \cite{hassani2018big}.  This section presents the key challenges and future directions related to blockchain big data research. The summary of research challenges upon the integration of blockchain and big data is described in Table \ref{tab:table1}.

\begin{table*}[h!]
\centering
\caption{Research challenges in integration of blockchain and big data.}
\label{tab:table1}
\resizebox{\textwidth}{!}{%
\begin{tabular}{|c|p{1.7cm}|p{3.75cm}|p{6.0 cm}|p{4.5cm}|}
\hline
\multicolumn{1}{|c|}{\textbf{Ref.}} &
  \multicolumn{1}{c|}{\textbf{Challenges}} &
  \multicolumn{1}{c|}{\textbf{Application}} &
  \multicolumn{1}{c|}{\textbf{Description}} &
  \multicolumn{1}{c|}{\textbf{Benefits}} \\ \hline
  \multirow{3}{*} {\cite{hassani2018big}} &
  \multirow{7}{*}{\makecell{Security and \\privacy \\ enhancement \\in big data}} &
  Big data and cryptocurrency. &
  Integration of big data and cryptocurrency for decentralized data management. &
  Secured data sharing and decentralized data access. \\ \cline{1-1} \cline{3-5} 
 \multirow{2}{*}{\cite{ liu2019survey1}} &
   &
  Blockchain smart contracts for big data. &
  Vulnerability scan and programming correctness for security and correctness in smart contracts' operations. &
  Secured data sharing and privacy. \\ \cline{1-1} \cline{3-5} 
 \multirow{2}{*}{\cite{ tariq2019security}} &
   &
  IoT big data, blockchain and fog computing. &
  Big data security in fog enabled IoT using blockchain. &
  Secured data transactions with low latency response. \\ \hline
 \multirow{3}{*}{\cite{akcora2018blockchain}} &
  \multirow{6}{*}{\makecell{Security \\and privacy in \\big data \\exchange}} &
  E-crime detection and bitcoin price predictions. &
  Interpretation of data stored in public blockchain. &
  Secured blockchain data transactions. \\ \cline{1-1} \cline{3-5} 
 \multirow{2}{*}{\cite{ yang2018smart}} &
   &
  Smart toy assisted with MEC and blockchain. &
  Smart contracts are used for validating various data exchanges authorized using blockchain. &
  Secured and low-latency response in smart toy business. \\ \cline{1-1} \cline{3-5} 
 \multirow{2}{*}{\cite{chen2019fade}} &
   &
  Big data exchange and smart contract. &
  A fair way for protecting user data copyright and ensures privacy using SC. &
  Privacy in decentralized big data sharing. \\ \hline
 \multirow{2}{*}{\cite{gramoli2018blockchain}} &
  \multirow{4}{*}{\makecell{Blockchain \\standardization}} &
  Blockchain and DLT. &
  High-level functional architecture for blockchain and DLT. &
  Standards for various functionalities in blockchain. \\ \cline{1-1} \cline{3-5} 
 \multirow{2}{*}{\cite{hofmann2017immutability}} &
   &
  Early standardization for blockchain immutability. &
  Describes different levels of standardization and their importance. &
  Participatory standard for blockchain immutability. \\ \hline
 \multirow{2}{*}{\cite{8768367}} &
  \multirow{5}{*}{\makecell{Complexity in \\big data}} &
  Crime big data. &
  Mining of data using various state-of-art data mining techniques. &
  Efficient mining of data for crime departments. \\ \cline{1-1} \cline{3-5} 
 \multirow{3}{*}{\cite{zheng2016big}} &
   &
  Big data in mobile network optimization. &
  Explorers the features of big data from the perspective of users and network operators.&
  Effective mobile data management. \\ \hline
 \multirow{2}{*}{\cite{ tan2020blockchain} } &
  \multirow{9}{*}{\makecell{Computational \\overhead in \\blockchain}} &
  Cyber physical social systems. &
  A lightweight blockchain for big data. &
  Privacy Preserving data transaction with low-latency. \\ \cline{1-1} \cline{3-5} 
 \multirow{2}{*}{\cite{mcconaghy2016bigchaindb}} &
   &
  Blockchain in distributed databases. &
  Blockchain on distributed databases allows 1-million writes for a second. &
  Scalability and faster querying with sub-second latency. \\ \cline{1-1} \cline{3-5} 
 \multirow{3}{*}{\cite{8327543}} &
   &
  Blockchain in cloud based healthcare big data. &
  Suggests off chain computation of healthcare data with control in the   blockchain. &
  Secured and immutable medical transactions. \\ \cline{1-1} \cline{3-5} 
  \multirow{2}{*}{\cite{8844768} }&
  &
  Blockchain for supply chains with 5G MEC. &
  One-way hash and bitwise rotation make the system light. &
  Low-latency  response. \\ \hline
  \multirow{2}{*}{\cite{9055743}} &
  \multirow{2}{*}{\makecell{Network \\Virtualization}} &
  SDN, big data, blockchain and 5G MEC. &
  SDN and big data are integrated with faster 5G and immutable blockchain. &
  Faster query processing and secured data transactions. \\ \hline
\end{tabular}%
}
\end{table*}

\subsection{{Research Challenges} }
\subsubsection{{Security in Blockchain}}
The blockchain, a valid ledger keeps track of various digital transactions across diversified domains such as IoT applications (includes data transaction from heterogeneous devices), in fifth-generation (5G) network, healthcare and financial services. Some of the notable services of the blockchain through decentralization are data security and privacy demanding more computational power (about 50\%) for the malicious users trying to deceive the block information \cite{chen2015resilient}. This type of attack is called 51\% attack. Though 51\% of the computational resources are required for any user to deceive information from the blocks, the double-spending attack is still possible. Blockchain smart contracts reinforce the environment to avoid the double-spend attacks \cite{liu2019survey1}. Blockchains' distributed nature (which shares each transaction network-wide) will induce greater complexity for fraudulent block transactions. Though blockchain and bigdata is a great marriage, data security issues concerned with big data and the data analytics models for handling big data must be considered. As the blockchain stores history of all transaction in the same state as it was performed makes it an essential candidate for big data application. Data-intensive applications like the healthcare industry where big data is employed for managing the voluminous data from medical practitioners, patients, clinicians, laboratory and pharmaceutical requires privacy-preserving data sharing. The researchers suggest that the secured blockchain framework can be employed for controlled access to the voluminous data using its decentralized data management \cite{ li2020survey}.  

Moreover, the permissionless blockchains which allow any user to join the chain without permission are secured by the hyper ledger \cite{ namasudra2020revolution}. Hyperledger strengthens the permissionless blockchain by allowing the users involved in the transaction to join through permissioned blockchain guaranteeing data provenance. One of the possible network threats to these kinds of a public blockchain is a Sybil attack, which enables a node in a blockchain to add enormous malicious users under its control \cite{ halpin2017introduction}. But, the PoW consensus algorithms will mitigate these attacks by allowing the malicious node to devote more of its computational resources to accomplish the attack.  An analysis in \cite{ tariq2019security} suggests that resource-constrained IoT environments where more data is accumulated from varied sources cannot be secured with traditional cryptographic infrastructure. Also, the authors recommend that the secured, distributed, and anonymized nature of blockchain is essential challenger for such environments. Furthermore, lightweight blockchain should be preferred for optimal computational resource utilization in the resource-constrained IoT environment with big data services.

The distributed ledger technology ensures trusted data transactions with immutability and transparency via peer-to-peer networking services. Blockchain assures better scalability than centralized architectures. But, as the chain grows longer and longer, the entries in the blockchain will be more and computational load in processing the data will increase tremendously. In blockchain applications like IoT, the nodes are simple and resource-constrained. Still, the security capability using cryptographic functions in blockchain consumes more computational resource for key exchange, encryption, decryption and digital signatures. The miners (the node that performs mining) which are responsible for creating new blocks and linking it with the existing chain requires higher computational load \cite{ salman2018security}. 

Therefore, the cryptographic techniques or security measures used for enhanced security in the blockchain environment should not impose more computational resources. The application of blockchain is increasing every day with an increase in the complexity of the data stored in the blockchain. Henceforth,  blockchain data analytics must be explored to ensure better performance of blockchain with varied complexity of the data \cite{ akcora2018blockchain}. Also, before integrating the blockchain with other technologies like big data, the type of blockchain public or private, security measures adopted, data processing capabilities should be considered for network safety and better performance.

\subsubsection{{Standardization}}
Blockchain was initially developed as the solution to the problem of digital cash (the cryptocurrency named bitcoin). It facilitates secure transaction of digital assets over different banks. Blockchain automized the global payment over the Internet irrespective of any topographical constraints within hours. Whereas the traditional financial system takes many days to perform any financial transactions worldwide.  Nevertheless,  the scope of adoption of blockchain has been hindered by its interoperability challenges. These challenges not only include the differences among different cryptocurrencies but also consists of the differences in the diversified transaction. Therefore, it is tedious for the blockchains to interoperate and integrate compatibly with the legacy systems. This, in turn, may hinder the regulatory acceptance of blockchains. One possible solution for this type of open systems is standardization to provide common technical guidelines for any industry.

An analysis of blockchain terminologies and various initiatives taken by non-profit organizations for standardization for blockchain was carried out in \cite{gramoli2018blockchain}. For standardizing distributed ledger technology (DLT) and blockchain, the international organization for standardization (ISO) which develops and publishes standards, has formed an ISO/TC 307 technical committee led by standards Australia for standardizing DLT and blockchain. The primary motive of this committee is to publish the standards related to blockchain privacy, taxonomy, smart contract, security (for users and data), privacy, interoperability, governance, and various use cases of blockchain. The different workgroups and their activities under ISO/TC 307 are summarized in \cite{cha2019international}. International telecommunication union (ITU), the working group under ISO, focuses on identifying and standardizing the DLT application, its services, best practise to be adopted for its implementation and further research on related standards. The world wide web consortium (W3C) which implements web standards has initiated standards for developing blockchain message formats (ISO20022), guidelines on blockchain storage (public, private and side-chain) and approving the use-cases. IEEE has developed a standard framework for blockchain use in IoT and a handbook on blockchain asset exchange. The Internet engineering task force (IETF), an open group that develops interoperability standards for network communications has a greater impact on blockchain standardization. 

Furthermore, a framework in \cite{hofmann2017immutability} was designed for the implementation of blockchain immutability. Also, they have discussed the effects of early standardization in blockchain immutability. They suggest that three different types of standards, namely, anticipatory, participatory, and responsive standards. The anticipatory standards are developed before the acceptance of a new service or technology. The participatory standard is developed and adopted during the implementation of the technology to test the conformance specification. And the responsive standards are adopted after technology adoption (or during its evolution). A framework for participatory standards to deploy the immutability concept of blockchain and its operation is discussed. 

Blockchain for big data allows the data sharing in cross-domain environment irrespective of the risk factors concerned with accumulating data from various data silos. Therefore, while adopting blockchain, proper standards and guidelines should adhere to the smooth functioning of the technology. 

\subsubsection{{Complexity of Big Data}}
The emergence of cloud computing, smart IoT applications have led to the massive accumulation of data. Along with the enormous growth of data in this information age, data management issues like inaccessible data, dirty or unclean data, and data privacy have also increased \cite{jin2015significance}. With the advent of big data, data quality management is more challenging. Furthermore, while handling more significant and complex datasets, the companies should ensure the authenticity of data source, cleanliness, and data breach. Because of this, the complete digital transformation of entire legacy data is still a challenging issue. The security perspective of the data management can be assured by blockchain, but yet, the complexity in big data management should be considered on its integration. 

The prominent challenges of big data are due to the nature of the data, conventional analysis models and inefficient data processing systems. The big data is inherently complex, making it challenging to represent and interpret, thereby increasing the computational complexity. The big data has heterogenic sources that exhibit different patterns and behaviors. Some of the essential characteristics are complex data type, its structure, more intricate relationships and wide-ranging quality. The big data mining activities such as data retrieval, analyzing the topic, text mining (sentiments and semantics extraction) will be challenging than the traditional data \cite{8768367}. The lack of knowledge of these characteristics and domain-specific data processing techniques will result in inefficient computational models.  A clear understanding of the attributes of inherently complex big data is mandatory for designing the computational models with the highest level of abstraction. Apart from diversified sources and massive volume, the critical feature of big data is its dynamically changing data (real-time information) \cite{kaisler2013big}. 

The big data processing systems are complex enough in handling the inherent complexity of the big data. These systems were built with high processing capability with more computational resource requirement. The system complexity includes the elaborate architecture, different processing modes and computing requirements. Basic knowledge of the system complexity will directly impact the performance of the big data systems. Also, the parameters affecting the energy utilization of big data processing systems must be considered while designing a robust framework. Some of them are system throughput, energy consumption, resource utilization, distributed data storage, parallel computation and accuracy in job calculation.

Furthermore, big data offers more chances for mobile networks to improve their service quality. The study in \cite{zheng2016big} has explored the integration of big data with mobile network optimization, with a focus on investigating the characteristics of big data from the perspective of the mobile network operator and users. The user-specific data obtained from user equipment include profile (location, communication pattern), behavior and other application data. The data of network operators include data from the core network, radio access network and Internet service providers. The core network provides data related to network performance, call details and application usage index. Information sourced from radio access network includes cell configuration, mobility, handover details, resource utilization (source details and link utilization), interference details, signal measurements and notification signal messages among different components in the mobile network. The effectiveness of the service laid by mobile network depends on how effectively network operators process this information and make valuable decisions. Efficient data analytic mechanisms are essential for better network optimization.

Therefore, the complex nature of the big data must be ensured while integrating it with blockchain as it will improve the way how the data is handled in big data processing models. Also, the mapping between complexity vs computation, energy consumption vs efficiency should be evaluated for laying out effective means of data sharing, trusted transactions, data access, intruder detection and enhanced security through decentralized blockchains.

\subsection{{Future Directions}} 
Big data which is proprietary for variability, volume, veracity, value and complexity, requires the data processing systems with higher computational capability. Also, the decentralized distributed ledger blockchain offers immutable, secured, and transparent data transactions that require more computational power for effective services. Upon integration of these complicated big data with blockchain, incur an unexpected computational complexity leading to the poor performance of the system. Therefore adaptive blockchain designs should be preferred, thereby alleviating the computational resource utilization for blockchain and 5G network communication can be utilized for faster services. This section presents the future directions for the integration of blockchain with big data.
\subsubsection{{Adaptive Blockchain Design for Big Data}}
The adaptive blockchain reduces the computational power required for processing the blocks even if the chain grows exponentially. The most preferred adaptive blockchain designs are lightweight blockchain for real-time big data and scalable blockchain for large scale big data.The framework in \cite{tan2020blockchain} was designed for large scale and real-time big data application cyber-physical social systems (which integrates cyber, physical and social systems)  uses blockchain for access control. The framework uses fog computing at the edge nodes for processing the local data dynamically. A lightweight symmetric algorithm is used for encryption for privacy-preserving data transactions. The cyber-physical social system big data is accessed using the account address of blockchain node. The access control details such as authentication, authorization are stored and managed in a blockchain. Experiments proved that the system is feasible and efficient, but it incurs more time to ensure privacy as all the authorizations are performed in the blockchain. Also, the retrieval mechanisms must be strengthened for better performance.

BigChainDB, a scalable blockchain for distributed big data was proposed in \cite{mcconaghy2016bigchaindb} by integrating the traditional blockchain with distributed databases to attain scalability through faster querying mechanisms. Blockchain network allows a user to join the chain based on the consensus. PoW and PoS are basic consensuses approach evolved to allow anyone to enter the network based on hash rate and stake(digital coins) respectively. BigChainDB can let 1-million writes for a second with sub-second latency (less than a second) and petabyte capacity. Also, it enforces permissioning system that can be employed for both public and private blockchains. Similarly, HBasechainDB framework in \cite{sahoo2018hbasechaindb}, a scalable blockchain for big data store deployed on the Hadoop ecosystem. The immutable and decentralization nature of the blockchain is built on the Hbase database of Hadoop. Instead of scaling the chain, the blockchain is implemented on a distributed database. Results proved that the system scales up linearly with sub-second latency and higher transaction throughputs. But, this framework is well suited for blockchain adoption to the organizations running on the Hadoop ecosystem.
Therefore, a standard adaptive framework supporting different types of blockchain for the big data system should be explored.

\subsubsection{{Blockchain for 5G Big Data}}
The diverse requirements change and an exponential increase in the number of mobile devices make it difficult for the 4G to meet future demands.  Though 5G has been integrated with technologies like SDN, Cloud, ML, network virtualization, it is still hard to meet the diverse requirements change. Also, these technologies procure different types of challenges in terms of decentralization, security and privacy, transparency, interoperability and immutability. Therefore, the blockchain with its feature will be an essential aspirant for massive computation with 5G big data applications \cite{9119406}. \textcolor{black}{In the 5G era, big data can rely on some key technologies to build its platforms such as  cloud computing, mobile edge computing, and software defined networking. In this context, blockchain can come as viable solution for realizing 5G technology-based big data services.}

% \begin{enumerate} 
%     \item \textbf{{Blockchain for cloud-based big data}} 

% \noindent 
\textbf{Blockchain for cloud-based big data}: 
A framework was proposed in \cite{zheng2019bcbim} to enhance the current building information modeling (BIM) by integrating tamper-resistant blockchain and mobile cloud through big data sharing.  BIM collects a huge amount of data throughout the project and needs to access the historical data for making certain decisions. The framework uses a private blockchain authorized by a trusted center for audit and data provenance in BIM as a cloud service (outsourced storage and computation). The system is evaluated based on the block size, security, hashing time and packaging period. The results show that blockchain can effectively resolve security issues and data quality with BIM and promotes it further. Likewise, a blockchain framework for cloud-based healthcare data handling was proposed in \cite{8327543}. Though blockchain ensures immutability and secured transaction of medical records stored in the cloud, the authors suggest off-chain storage for healthcare data due to protection of privacy laws and dynamic changing behavior of healthcare data (old record might be unusable). But, the immutable hashes can be stored on the chain, whereas off-chain data can be modified saved as a distributed database.
% \item \textbf{{Blockchain for mobile edge computing based big data}}
% \noindent \textbf{{Blockchain for mobile edge computing based big data}:

% \noindent 
\textbf{Blockchain for mobile edge computing based big data}:
The integration of mobile edge computing (MEC) and blockchain is a right solution to achieve low latency response which shrinks the computing power by local data processing. A distributed blockchain framework was proposed in \cite{8951115} for privacy protection with heterogeneous MEC in 5G and beyond networks. The consensus in multidomain collaborative routing is achieved without exposing the network topology. Blockchain is used for multilevel mutual trust and collaborative routing. The privacy of heterogeneous MEC collaboration in 5G is highly improved with blockchain adoption. 

 A prototype for a smart toy, an IoT device based on edge computing and blockchain (hyper ledger fabric) was proposed in \cite{ yang2018smart} for secured data exchange.  Smart toy environment has many different types of data exchanges among single supplier, single demander, multiple supplier and multiple demanders. For each data exchange, the ids are stored and validated through smart contracts (handles accounting). Chaincode, a blockchain-based smart contract handles the complex accounting in the smart toy business. The edge computing is used for local client computations ensuring low- latency response. The framework ensures a secured, flexible, scalable and confidential data exchange among smart toy business participants. Similarly, a lightweight blockchain with RFID enabled authentication system for supply chains with 5G MEC was proposed in \cite{8844768}. The one-way hash function and bitwise rotation make the system light with less computational power compared to existing protocols. 

% \item \textbf{Blockchain for software defined networking-based big data}
% \noindent 
\textbf{Blockchain for software defined networking-based big data}:
Software defined networking (SDN) makes the network agile and flexible by improving the network control with faster response for changing requirements. Upon integration of big data and SDN, both the technologies are benefited seamlessly \cite{7389832}. SDN can assist big data in solving most of its issues like data delivery, data processing in the cloud, transmission, data scheduling and resource optimization. Likewise, big data can assist SDN  in handing traffic data, security vulnerabilities and inter and intra data center network communications. Though they serve each other effectively, there are some open issues to be resolved. The overflooded queries in the SDN controller may degrade its performance, in turn,  making it insufficient for accommodating more extensive big data entries, thereby imposing the scalability challenge and central point of failure. Resource-constrained and unintelligent network switches can flood the SDN controller with raw data packets imposing a higher computational overhead. Also there is no high-level programming environment available for the development of big data applications. Above all, security vulnerabilities are more in SDN. Therefore the SDN and big data can serve each other better when integrated with faster 5G and the scalable, immutable and decentralized blockchain. The 5G enables faster processing speed, whereas blockchain can resolve scalability and security issues in SDN \cite{9055743}. The fusion of SDN with blockchain and MEC \cite{8875714} will solve the majority of issues in network virtualization, making it fit for processing big data applications.
% \end{enumerate}

Therefore, big data will be promoted when the blockchain technology meets MEC, SDN, Cloud with robust 5G communications. All these technologies together ensure security and privacy, scalability, parallel computing, transparency and trust when individual challenges are met for compatible functioning of the underlying technologies.

\section{{Conclusions}} 
\label{Sec:Conclusion}
Blockchain is a disruptive ledger technology that has sparked a significant interest to support big data systems with high security and efficient network management. In this article, we have conducted a state-of-the-art review on the application of blockchain for big data. We have first discussed the recent advances in blockchain and big data and explained the motivation behind the integration of these two technologies. Particularly, we have provided an extensive survey on the use of blockchain in a number of key big data services, including big data acquisition, big data storage, big data analytics, and big data privacy preservation. Then, we have explored the opportunities brought by blockchain in important big data applications, such as smart city, smart healthcare, smart transportation, and smart grid. The emerging blockchain-big data platforms and projects have been also highlighted and analyzed. From the extensive literature review on blockchain-big data services and applications, we have identified some key technical challenges and pointed out possible future directions to spur further research in this promising area.

Therefore, it is evident from the discussion that, a robust blockchain framework for bigdata encompasses numerous technical challenges to be considered upon its integration and deployment. The challenges pertaining to the blockchain can be resolved with bigdata and the vice versa. The substantial data accumulations and data services of the big data can be effectively managed and secured by the blockchain. The decentralized and immutable ledger with advanced technologies ensures data integrity and bigdata analytics provides better insights for making valuable predictions for massive data accumulation.

\bibliographystyle{IEEEtran}
\bibliography{Bibliography}
\end{document}